\documentclass[twocolumn,showpacs,prl]{revtex4}
\usepackage{amsmath,amssymb,graphicx,color}
\usepackage{color}

\begin{document}

\title{Exploring the thermodynamic limit of Hamiltonian models: convergence to the Vlasov equation.}

\author{Andrea Antoniazzi$^{1}$\thanks{andrea.antoniazzi@unifi.it},
Francesco Califano$^
{2}$\thanks{francesco.califano@df.unipi.it}, 
Duccio Fanelli$^{1,3}$\thanks{duccio.fanelli@ki.se}, 
Stefano Ruffo$^{1}$\thanks{stefano.ruffo@unifi.it}}

\affiliation{ 1. Dipartimento di Energetica and CSDC, Universit\`a di
  Firenze, and INFN, via S. Marta, 3, 50139 Firenze, Italy\\
  2. Dipartimento di Fisica "E.Fermi" and CNISM, Universit\`a di
  Pisa, Largo Pontecorvo, 3 56127 Pisa, Italy\\
  3. Theoretical Physics, School of Physics and Astronomy, University of Manchester, Manchester M13 9PL, United Kingdom} \date{\today}

\begin{abstract}
We here discuss the emergence of Quasi Stationary States (QSS), a universal feature of systems 
with long-range interactions. With reference to the Hamiltonian Mean
Field (HMF) model, numerical simulations are performed based on both the original $N$-body
setting and the continuum Vlasov model which is supposed to hold in the thermodynamic 
limit. A detailed comparison unambiguously demonstrates that the Vlasov-wave system provides the correct 
framework to address the study of QSS. Further, analytical calculations based on Lynden-Bell's
theory of violent relaxation are shown to result in accurate predictions. Finally, in specific regions of parameters space, 
Vlasov numerical solutions are shown to be affected by small scale fluctuations, a finding that 
points to the need for novel schemes able to account for particles correlations.  
\end{abstract}


\maketitle

The Vlasov equation constitutes a universal theoretical framework and plays a role of paramount 
importance in many branches of applied and fundamental physics. Structure formation in the 
universe is for instance a rich and fascinating problem of classical physics: The fossile radiation 
that permeates the cosmos is a relic of microfluctuation in the matter created by the Big Bang, and 
such a small perturbation is believed to have evolved via gravitational instability to the pronounced 
agglomerations that we see nowdays on the galaxy cluster scale. Within this scenario, gravity is hence 
the engine of growth and the Vlasov equation governs the dynamics of the non baryonic ``dark matter" \cite{peebles}. 
Furthermore, the continuous Vlasov description is the reference model for several space and laboratory plasma applications, 
including many interesting regimes, among which the interpretation of coherent electrostatic structures 
observed in plasmas far from thermodynamic equilibrium. The Vlasov equation is obtained as the mean--field 
limit of the $N$--body Liouville equation, assuming that each particle interacts with an average field generated by all 
plasma particles (i.e. the mean electromagnetic field determined by the 
Poisson or Maxwell equations where the charge and current densities are 
calculated from the particle distribution function) while inter--particle 
correlations are completely neglected.

%Finally, semiclassical Vlasov
%equation and the Feynman path integral method are being used as an effective many-body description of nuclear
%dynamics below the Coulomb barrier \cite{bonasera}. 

Numerical simulations are presently one of the most powerful resource 
to address the study of the Vlasov equation.
%such physical systems and represent a major
%challenge in computational physics. 
In the plasma context, the Lagrangian
Particle-In-Cell approach is by far the most popular, while 
Eulerian Vlasov codes are particularly suited for 
analyzing specific model problems, due to the associated low 
noise level which is secured even in the non--linear 
regime \cite{mangeney}. However, any numerical scheme designed to integrate the 
continuous Vlasov system involves a 
discretization over a finite mesh. This is indeed an  
unavoidable step which in turn affects numerical accuracy.  
A numerical (diffusive and dispersive) characteristic length is in fact introduced 
being at best comparable with the grid mesh size: as soon as the latter matches the 
typical length scale relative to the (dynamically generated) fluctuations 
a violation of the continuous Hamiltonian character of the equations occurs 
(see Refs. \cite{califano}). It is important to emphasize that
even if such {\em non Vlasov} effects are strongly localized 
(in phase space), the induced large scale topological changes will 
eventually affect the system globally. Therefore, aiming at clarifying the 
problem of the validity of Vlasov numerical models, it is crucial to compare
a continuous Vlasov, but numerically discretized, approach 
to a homologous N-body model.  

Vlasov equation has been also invoked as a reference model in many interesting one dimensional problems,
and recurrently applied to the study of wave-particles interacting systems. The Hamiltonian Mean Field 
(HMF) model~\cite{antoni-95}, describing the coupled motion of $N$ rotators, is in particular assimilated to 
a Vlasov dynamics in the thermodynamic limit on the basis of rigorous results~\cite{BraunHepp}. 
The HMF model has been historically introduced as representing 
gravitational and charged sheet models and is quite extensively analyzed as a paradigmatic 
representative of the broader class of systems with long-range interactions~\cite{Houches02}. 
%This system displays a gallery of peculiar behaviours ~\cite{Houches02} including {\it inequivalence} between microcanonical 
%and canonical ensemble, physically manifested through negative specific heat and temperature jumps,
%{\it broken ergodicity} and the existence of  {\it metastable states}.  
A peculiar feature of the HMF model, shared also by other long-range interacting systems, is the presence of
{\it Quasi Stationary States} (QSS). During time evolution, the system gets trapped in such states, which
are characterized by non Gaussian velocity distributions, before relaxing to the final Boltzmann-Gibbs 
equilibrium \cite{ruffo_rapisarda}. An attempt has been made \cite{rapisarda_tsallis} to interpret the 
emergence of QSSs by invoking Tsallis statistics \cite{Tsallis}. This approach has been later on 
criticized in \cite{Yamaguchi}, where QSSs were shown to
correspond to stationary stable solutions of the Vlasov equation, for a particular choice of the initial condition.  
More recently, an approximate analytical theory, based on the Vlasov equation, which derives the QSSs of the HMF model 
using a maximum entropy principle, was developed in ~\cite{antoniazziPRL}. This theory 
is inspired by the pioneering work of Lynden-Bell ~\cite{LyndenBell68} and relies on previous work
on 2D turbulence by Chavanis \cite{chava2D}. However, the underlying 
Vlasov ansatz has not been directly examined and it is recently being debated \cite{EPN}. 

In this Letter, we shall discuss numerical simulations of the continuous Vlasov model, the kinetic counterpart
of the discrete HMF model. By comparing these results to both direct N-body simulations and analytical predictions, 
we shall reach the following conclusions:
(i)  the Vlasov formulation is indeed ruling the dynamics of the QSS; (ii)  the proposed 
analytical treatment of the Vlasov equation is surprisingly accurate, despite the approximations involved in the
derivation;  (iii) Vlasov simulations are to be handled with extreme caution when exploring 
specific regions of the parameters space. 

The HMF model is characterized by the following Hamiltonian
\begin{equation}
\label{eq:ham}
H = \frac{1}{2} \sum_{j=1}^N p_j^2 + \frac{1}{2 N} \sum_{i,j=1}^N 
\left[1 -  \cos(\theta_j-\theta_i) \right]
\end{equation}
where $\theta_j$ represents the orientation of the $j$-th rotor and
$p_j$ is its conjugate momentum. To monitor the evolution of the
system, it is customary to introduce the magnetization, a macroscopic order
parameter defined as $M=|{\mathbf M}|=|\sum {\mathbf m_i}| /N$, where
${\mathbf m_i}=(\cos \theta_i,\sin \theta_i)$ stands for the microscopic
magnetization vector. As previously reported \cite{antoni-95}, after an initial transient, 
the system gets trapped into Quasi-Stationary States (QSSs), i.e. 
non-equilibrium dynamical regimes  whose lifetime diverges when increasing 
the number of particles $N$. Importantly, when performing the mean-field limit ($N
\rightarrow \infty$) {\it before} the infinite time limit, the system
cannot relax towards Boltzmann--Gibbs equilibrium and remains
permanently confined in the intermediate QSSs. 
%In other words, 
%the latter manifest because the infinite time and 
%the thermodynamic limits do not commute.
As mentioned above, this phenomenology is widely observed for systems with long-range
interactions, including galaxy dynamics~\cite{Padmanabhan}, free electron lasers~\cite{Barre},
2D electron plasmas~\cite{kawahara}.

In the $N \to \infty$ limit
%i.e. when the system is indefinitely stuck in the QSSs, 
the discrete HMF dynamics reduces to the Vlasov equation
\begin{equation}
\partial f / \partial t + p \, \partial f / \partial \theta \,\, 
- (dV / d \theta ) \, \partial f / \partial p = 0 \, , 
%\frac{\partial f}{\partial t} + p\frac{\partial f}{\partial \theta} -
%\frac{d V}{d \theta} \frac{\partial f}{\partial p}=0\quad ,
\label{eq:VlasovHMF}
\end{equation}
where $f(\theta,p,t)$ is the microscopic one-particle
distribution function and
\begin{eqnarray}
V(\theta)[f] &=& 1 - M_x[f] \cos(\theta) - M_y[f] \sin(\theta) ~, \\
M_x[f] &=& \int_{-\pi}^{\pi} \int_{-\infty}^{\infty}  f(\theta,p,t) \, \cos{\theta}  {\mathrm d}\theta
{\mathrm d}p\quad , \\
M_y[f] &=& \int_{-\pi}^{\pi} \int_{\infty}^{\infty}  f(\theta,p,t) \, \sin{\theta}{\mathrm d}\theta
{\mathrm d}p\quad .
\label{eq:pot_magn}
\end{eqnarray}
The specific energy $h[f]=\int \int (p^2/{2}) f(\theta,p,t) {\mathrm d}\theta
{\mathrm d}p - ({M_x^2+M_y^2 - 1})/{2}$ and momentum
$P[f]=\int \int p f(\theta,p,t) {\mathrm d}\theta
{\mathrm d}p$ functionals are conserved quantities. Homogeneous states are characterized by $M=0$, 
while non-homogeneous states correspond to $M \ne 0$. 

Rigorous mathematical results \cite{BraunHepp} demonstrate that, indeed, the Vlasov framework applies in the continuum 
description of mean-field type models. This observation corroborates the claim that any 
theoretical attempt to characterize the QSSs should resort to the above Vlasov based 
interpretative picture. Despite this, the QSS non-Gaussian velocity
distributions have been {\em fitted}~\cite{rapisarda_tsallis} using Tsallis' $q$--exponentials,
and the Vlasov formalism assumed valid {\em only} for the limiting case of homogeneous initial 
conditions \cite{EPN}. In a recent paper \cite{antoniazziPRL}, the aforementioned 
velocity distribution functions were instead reproduced with an analytical expression derived from the Vlasov scenario, 
with no adjustable parameters and for a large class of initial conditions, including inhomogeneous ones. 
The key idea dates back to the seminal work by Lynden-Bell \cite{LyndenBell68} (see also \cite{Chavanis06},
\cite{Michel94}) and consists in coarse-graining the microscopic one-particle distribution function
$f(\theta,p,t)$ by introducing a local average in phase space. It is then possible to
associate an entropy to the coarse-grained distribution $\bar{f}$: The corresponding statistical equilibrium
is hence determined by maximizing such an entropy, while imposing the conservation of the Vlasov dynamical invariants, 
namely energy, momentum and norm of the distribution. We shall here limit our discussion to the case 
of an initial single particle distribution which takes only two distinct values: $f_0=1/(4
\Delta_{\theta} \Delta_{p})$, if the angles (velocities) lie within an
interval centered around zero and of half-width $\Delta_{\theta}$
($\Delta_{p}$), and zero otherwise. This choice corresponds to the
so-called ``water-bag" distribution which is fully specified by energy
$h[f]=e$, momentum $P[f]=\sigma$ and the initial magnetization ${\mathbf
M_0}=(M_{x0}, M_{y0})$.  The maximum entropy calculation is then performed analytically \cite{antoniazziPRL}
and results in the following form of the QSS distribution 
%%%%%%%%%%%%%%%%%%%%%%%%%%%%%%%%%%%%%%%%%%%%%%%%%%%%%%%%%%%%%%%%
\begin{equation}
\label{eq:barf} \bar{f}(\theta,p)= f_0\frac{e^{-\beta (p^2/2
- M_y[\bar{f}]\sin\theta
- M_x[\bar{f}]\cos\theta)-\lambda p-\mu}}
{1+e^{-\beta (p^2/2  - M_y[\bar{f}]\sin\theta
 - M_x[\bar{f}]\cos\theta)-\lambda p-\mu}}
\end{equation}
%%%%%%%%%%%%%%%%%%%%%%%%%%%%%%%%%%%%%%%%%%%%%%%%%%%%%%%%%%%%%%%%
where $\beta/f_0$, $\lambda/f_0$ and $\mu/f_0$ are rescaled Lagrange multipliers, respectively associated to the
energy, momentum and normalization. Inserting expression (\ref{eq:barf}) into the above 
constraints and recalling the definition of $M_x[\bar{f}]$, $M_y[\bar{f}]$, one obtains an implicit system 
which can be solved numerically to determine the Lagrange multipliers and the expected magnetization in the QSS. Note that the
distribution (\ref{eq:barf}) differs from the usual Boltzmann-Gibbs expression because of the
``fermionic'' denominator. Numerically computed velocity distributions have been compared in \cite{antoniazziPRL} 
to the above theoretical predictions (where no free parameter is used), obtaining an overall good agreement.
%Although not a single free parameter was used, an excellent overall agreement was found in the tails of the
%distribution. 
However, the central part of the distributions is modulated by the presence of two symmetric bumps, which 
are the signature of a collective dynamical phenomenon \cite{antoniazziPRL}. The presence of these bumps is not 
explained by our theory. Such discrepancies has been recently claimed to be an indirect proof of the fact 
that the Vlasov model holds only approximately true.
%with reference 
%to the specific HMF case, thus justifying the search for novel theoretical frameworks, alternative to conventional 
%statistical mechanics. 
We shall here demonstrate that this claim is not correct and that the deviations between theory 
and numerical observation are uniquely due to the approximations built in the Lynden-Bell approach. 

A detailed analysis of the Lynden-Bell equilibrium (\ref{eq:barf}) in the parameter plane 
$(M_{0},e)$ enabled us to unravel a rich phenomenology, including out of equilibrium phase transitions 
between homogeneous ($M_{QSS}=0$)  and non-homogeneous ($M_{QSS} \ne 0$) QSS states. 
Second and first order transition lines are found that separate homogeneous and non homogeneous states and
merge into a tricritical  point approximately located in $(M_{0},e)=(0.2,0.61)$. When the transition is 
second order two extrema of the Lynden-Bell entropy are identified in the inhomogeneous phase: the solution  
$M=0$ corresponds to a saddle point, being therefore unstable; the global maximum is 
instead associated to $M \neq 0$, which represents the equilibrium predicted by the theory. This argument is
important for what will be discussed in the following.

\begin{figure}[htbp]
  \centering
%  \vspace*{2.5em}
  \includegraphics[width=7cm]{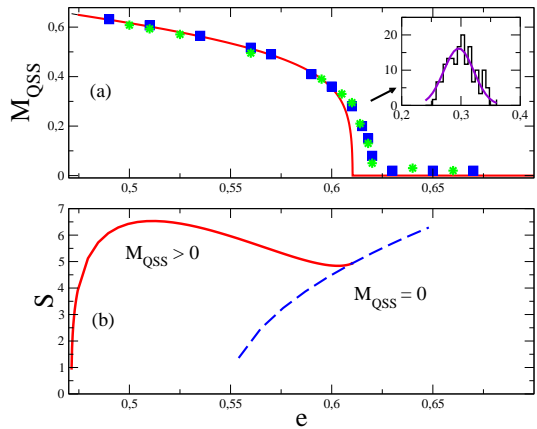}
  \caption{Panel (a): The magnetization in the QSS is plotted as function of energy, $e$, at $M_0=0.24$. 
  The solid line refers to the Lynden-Bell inspired theory. Stars (resp. squares) stand for Vlasov 
  (resp. N-body) simulations. Inset: Probability distribution of  
  $M_{QSS}$ computed via N-body simulation (the solid line is a Gaussian fit).    
  Panel (b): Entropy $S$ at the stationary points, as function of energy, $e$: 
  magnetized solution (solid line) and non--magnetized one (dashed line).}
  \label{fig1}
\end{figure}

Let us now turn to direct simulations, with the aim of testing the above scenario, and focus first on the kinetic model 
(\ref{eq:VlasovHMF})--(\ref{eq:pot_magn}). The algorithm solves the Vlasov equation in phase space and uses the 
so-called ``splitting scheme", a widely adopted strategy in numerical fluid dynamics. Such a scheme, pioneered by
Cheng and Knorr \cite{Cheng}, was first applied to the study of the Vlasov-Poisson equations in the electrostatic 
limit and then employed for a wide spectrum of problems \cite{califano}. For different values of the pair 
$(M_{0},e)$, which sets the widths of the initial water-bag profile, we performed a direct integration of 
the Vlasov system  (\ref{eq:VlasovHMF})--(\ref{eq:pot_magn}). After a transient, magnetization is 
shown to eventually attain a constant value, which corresponds to the QSS value observed in
the HMF, discrete, framework. The asymptotic magnetizations are hence recorded when varying the 
initial condition. Results (stars) are reported in figure \ref{fig1}(a) where $M_{QSS}$ is plotted as function of $e$. 
A comparison is drawn with the predictions of our theory (solid line) and with the outcome of N-body simulation (squares) 
based on the Hamiltonian (\ref{eq:ham}), finding an excellent agreement. 
This observation enables us to conclude that (i) the Vlasov equation governs the HMF dynamics for 
$N \to \infty$ {\it both} in the homogeneous and non homogeneous case; (ii) Lynden-Bell's violent relaxation 
theory allows for reliable predictions, including the transition from magnetized to non-magnetized states.  

Deviations from the theory are detected near the transition. 
This fact has a natural explaination and raises a number of fundamental questions related to the use of 
Vlasov simulations. As confirmed by the inspection of figure \ref{fig1}(b), close to the transition point, 
the entropy $S$ of the Lynden-Bell coarse-grained distribution takes almost the same value when evaluated 
on the global maximum (solid line) or on the saddle point (dashed line). The entropy 
is hence substantially flat in this region,  which in turn implies that there exists an extended basin 
of states accessible to the system. This interpretation is further validated by the inset of figure \ref{fig1}(a), 
where we show the probability distribution of $M_{QSS}$ computed via N-body simulation. The bell-shaped profile 
presents a clear peak, approximately close to the value predicted by our theory. Quite remarkably, 
the system can converge to final magnetizations which are sensibly different from the expected value. 
Simulations based on the Vlasov code running at different resolutions (grid points) confirmed 
this scenario, highlighting a similar degree of variability. These findings point to the fact that in 
specific regions of the parameter space, Vlasov numerics needs to be carefully analyzed (see also Ref.~\cite{Elskens}). 
Importantly, it is becoming nowadays crucial to step towards an ``extended«« Vlasov theoretical model which enables 
to account for discreteness effects, by incorporating at least two particles correlations interaction term. 

%%%%%%%%%%%%%%%%%%%%%%%%%%%%%%%%%%%%%%%%%%%%%%%%%%%%%%%%%%%%%%
%%%%%%%%%%%%%%%%%%%%%%%%%%%%%%%%%%%%%%%%%%%%%%%%%%%%%%%%%%%%%%
%%%%%%%%%%%%%%%%%%%%%%%%%%%%%%%%%%%%%%%%%%%%%%%%%%%%%%%%%%%%%%
\begin{figure}[htbp]
\centering
\includegraphics[width=7cm]{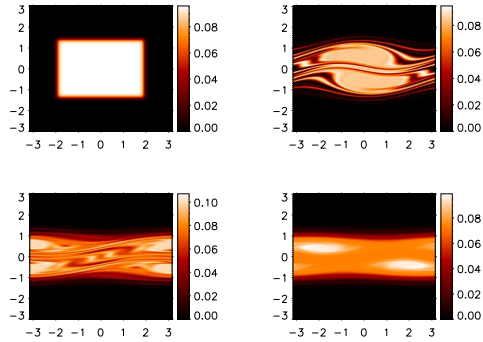}
%  \vspace*{2.5em}
\caption{Phase space snapshots for $(M_{0},e)=(0.5,0.69)$.}
\label{fig2}
\end{figure}
\begin{figure}[htbp]
  \centering
%  \vspace*{2.0em}
  \includegraphics[width=7cm]{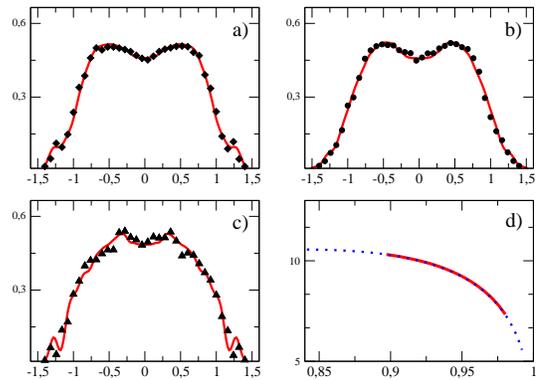}
  \vspace*{2.0em}
  \caption{Symbols: velocity distributions computed via N-body simulations. Solid line: velocity distributions 
  obtained through a direct integration of the Vlasov equation. Here $e=0.69$ and $M_0=0.3$ (a), $M_0=0.5$ (b), 
  $M_0=0.7$ (c). Panel (d): Entropy at the stationary points as a function of the initial magnetization: 
  the solid line refers to the global maximum, while the dotted line to the saddle point.}
  \label{fig3}
\end{figure}

%%%%%%%%%%%%%%%%%%%%%%%%%%%%%%%%%%%%%%%%%%%%%%%%%%%%%%%%%%%%%%
Qualitatively, one can track the evolution of the system in phase space, both for the homogeneous and non 
homogeneous cases. Results of the Vlasov integration are displayed in figure \ref{fig2} for  $(M_{0},e)=(0.5,0.69)$,  
where the system is shown to evolve towards a non magnetized QSS. The initial water-bag 
distribution splits into two large resonances, which persist asymptotically: the latter 
acquire constant opposite velocities which are maintained during the subsequent time evolution, in agreement with the findings of 
\cite{antoniazziPRL}. The two bumps are therefore an emergent property of the model, which is correctly 
reproduced by the Vlasov dynamics. For larger values of the initial magnetization ($M_{0}>0.89$), while keeping $e=0.69$, the
system evolves towards an asymptotic magnetized state, in agreement with the theory. In this case 
several resonances are rapidly developed which eventually coaelesce giving rise to complex patterns in phase space. 
More quantitatively, one can compare the velocity distributions resulting from, 
respectively, Vlasov and N-body simulations. The curves are diplayed in figure \ref{fig3} (a),(b),(c) for various 
choices of the initial conditions in the magnetized region. The agreement is excellent, thus reinforcing our former 
conclusion about the validity of the Vlasov model. 
Finally, let us stress that, when $e=0.69$, the two solutions (resp. magnetized and non magnetized)  
\cite{antoniazziPRL} are associated to a practically indistinguishible entropy level 
(see figure \ref{fig3} (d)). As previously discussed, the system explores an almost flat entropy landscape 
and can be therefore be stuck in local traps, because of finite size effects. A pronounced variability 
of the measured $M_{QSS}$ is therefore to be expected. 
%This observation contributes to 
%definitely solve the puzzle about the pecularity of the regime $e=0.69$, a reference value for the energy that has 
%attracted the interests of theoreticians during the last decade \cite{Houches02}. 
%\vspace*{2.0em}
%%%%%%%%%%%%%%%%%%%%%%%%%%%%%%%%%%%%%%%%%%%%%%%%%%%%%%%%%%%%%%%%%%%%5
%%%%%%%%%%%%%%%%%%%%%%%%%%%%%%%%%%%%%%%%%%%%%%%%%%%%%%%%%%%%%%%%%%%%

In this Letter, we have analyzed the emergence of QSS, a universal feature that occurs in systems with long-range
interactions, for the specific case of the HMF model. By comparing numerical simulations and analytical predictions, 
we have been able to unambiguously demonstrate that the Vlasov model provides an accurate framework to address 
the study of the QSS. Working within the Vlasov context one can develop a fully predictive theoretical approach, which is
completely justified from first principles. Finally, and most important, results of conventional Vlasov codes are to be
critically scrutinized, especially in specific regions of parameters space close to transitions from homogeneous
to non homogeneous states.

We acknowledge financial support from the PRIN05-MIUR project {\it Dynamics and thermodynamics of systems with long-range 
interactions}.
\vspace*{-0.6cm}
%%%%%%%%%%%%%%%%%%%%%%%%%%%%%%%%%%%%%%%%%%%%%%%%%%%%%%%%%%%%%%%%%%


\begin{thebibliography}{99}
\vspace*{-0.3cm}
\bibitem{peebles} P.J. Peebles, \emph{The Large-scale structure of the Universe}, Princeton, NJ: Princeton
Universeity Press (1980).
%\bibitem{bonasera} A. Bonasera et al., J. Phys. {\bf 23} 1297-1302 (1997) 
\bibitem{mangeney} A. Mangeney et al., J. Comp. Physics {\bf 179}, 495 (2002).
\bibitem{califano} L. Galeotti et al., Phys. Rev. Lett. {\bf 95}, 
015002 (2005); F. Califano et al., Phys. Plasmas {\bf 13}, 082102 (2006). 
\bibitem{antoni-95} M.~Antoni et al., Phys.~Rev.~E \textbf{52},
  2361 (1995).
\bibitem{BraunHepp} W. Braun et al., Comm.  Math. Phys. \textbf{56}, 101 (1977).
 \bibitem{Houches02} T.~Dauxois et al., Lect. Not. Phys. 
{\bf 602}, Springer (2002).
\bibitem{ruffo_rapisarda} V. Latora et al. Phys. Rev. Lett. \textbf{80}, 692 (1998).
\bibitem{rapisarda_tsallis} V. Latora et al. Phys. Rev. E \textbf{64} 056134 (2001).
\bibitem{Tsallis} C. Tsallis, J. Stat. Phys. {\bf 52}, 479 (1988).
\bibitem{Yamaguchi} Y.Y. Yamaguchi et al. Physica A, {\bf 337}, 36 (2004).
\bibitem{antoniazziPRL} A.~Antoniazzi et al., Phys. Rev. E \textbf{75} 011112 (2007); 
P.H. Chavanis Eur. Phys. J. B {\bf 53}, 487 (2006).
\bibitem{LyndenBell68} D. Lynden-Bell et al., Mon. Not. R. Astron. Soc.
{\bf 138}, 495 (1968).
\bibitem{chava2D} P.H. Chavanis, Ph. D Thesis, ENS Lyon (1996).
\bibitem{EPN} A. Rapisarda et al., Europhysics News, {\bf 36}, 202 (2005); 
F. Bouchet et al., Europhysics News, {\bf 37}, 2, 9-10 (2006).
\bibitem{Padmanabhan} T. Padmanabhan, Phys. Rep. {\bf 188}, 285 (1990).
\bibitem{Barre} J. Barr{\'e} et al., Phys. Rev E {\bf 69}, 045501(R) (2004).
\bibitem{kawahara} R. Kawahara and H. Nakanishi, cond-mat/0611694.
\bibitem{Chavanis06}  P.~H.~Chavanis, {Physica A} \textbf{359}, 177 
(2006).
\bibitem{Michel94} J. Michel et al., Comm. Math. Phys. 
\textbf{159}, 195 (1994).
\bibitem{Cheng} C.G. Cheng and G. Knorr, J. Comp. Phys. {\bf 22}, 330 (1976).
\bibitem{Elskens} M.C. Firpo, Y. Elskens, Phys. Rev. Lett. {\bf 84}, 3318 (2000).
\end{thebibliography}
\end{document}